# Mach's relativity of rotation in light of contemporary physics

Herbert I. Hartman

*William Rainey Harper College, Palatine, Illinois 60067*

Charles Nissim-Sabat*

*Northeastern Illinois University, Chicago, Illinois 60625,*

*and Cherskov and Flaynik, Chicago, Illinois 60606*

Mach argued for a relational rather than an absolute notion of space, insisting that centrifugal forces inside a rotating object such as a bucket can be reproduced by keeping the bucket fixed and rotating the universe. In response to a paper of ours denying the validity of Mach's views, Bhadra and Das elaborate on Mach's position. We address several of their arguments and show that Mach's relational notion of space is wrong-headed. Special and general relativity distinguish between a "bucket" (i.e. any system) rotating in a fixed universe and a bucket fixed in a rotating universe and between a non-rotating bucket in a non-rotating universe and a co-rotating bucket in a rotating universe, distinctions that go against Mach's relational theory of space. Even when taken on its own terms, Mach's theory can apply only to single point-like buckets rotating at infinitesimal angular velocities.

*e-mail: cnissim@hotmail.com

# I. Introduction

The long-standing debate on whether there is such a thing as absolute space often centers on the interpretation of Newton's famous rotating bucket experiment, that is, on what causes the centrifugal force on the rotating liquid that is observed in the bucket frame of reference. Newton maintained it is acceleration of the liquid with respect to absolute space.[1] Huygens, Leibniz, Berkeley, and most notably Mach have argued for a relational theory of space wherein the centrifugal force on the liquid is due to the acceleration of the liquid with respect to the masses of the universe. Mach maintained the relativity of all motion, including rotational motion,[2] claiming that one cannot distinguish between a bucket rotating in a fixed universe (BR) and the universe rotating around a fixed bucket (UR). Mach influenced Einstein, but most physicists agree that general relativity (GR) is not a Machian theory.

We have argued that Mach's theory of the relativity of rotational motion suffers from internal inconsistencies leading to the inherent contradiction that there cannot be a fixed bucket in a rotating universe, especially if frame-dragging is invoked.[3] We have also claimed that non-inertial experiments such as electro-magnetic and stellar aberration observations could distinguish between BR and UR situations. We discussed Mach's ideas in light of classical physics, the results of which he accepted but failed to fully take into account.

In response to Ref. 3, Bhadra and Das (BD) have discussed Mach's assumptions and their consequences.[4] We contend here that BD have not shown that it is possible to formulate a theory in which the BR and UR situations are indistinguishable. We focus on situations where an observer (universe-bound or, especially, bucket-bound) makes different observations in a UR and a BR situation. Only one such observation is needed to falsify Mach. In Sec. IIA we show that Mach cannot explain simple magnetic observations and in Sec. IV.E we show that he cannot explain the trajectory of a droplet ejected from a bucket in a UR situation. We offer several other examples where a Machian account is inadequate. While many of our arguments here are based on relativistic physics, we point out several more instances where classical physics invalidates Mach, raising anew the question why he believed his relational view of space was valid.

We use "bucket" or the letter B to denote any system possessing angular momentum (be it an elementary particle or a cluster of galaxies). Unless otherwise indicated, all motion is in the *x,y* plane, *r* denotes the distance from B's axis, *a* is B's radius, and *ω* is the relative angular velocity between B and the universe. "Rotation" denotes motion of an object or system around an internal axis and "revolution" denotes motion around an external axis.

We follow BD's order in Ref. 4, but make no attempt to address each of their arguments. In Sec. II we show that, in practice, Mach's theory of the relativity of all motions is restricted to the proposition that the rotating bucket and the rotating universe situations are indistinguishable, and show this proposition is inconsistent with classical physics and both special relativity (SR) and GR. In Sec. III we critique BD's notion of what can be "fixed" in a UR situation and add support to our earlier conclusion[3] that there are inherent contradictions in Mach's thinking. In Sec. IV we show that several non-inertial effects allow the detection of differences between BR and UR situations. BD's arguments concerning the equivalence of heliocentric and geocentric systems are discussed briefly in Sec. V. We summarize our conclusions in Sec. VI.

## II. The inadequacy of Mach's principle of the relativity of all motions.

In Ref. 4 BD point out that Mach maintained that all motions, including rotational motions, are the relative motions of material bodies: "*All* masses and *all* velocities, and consequently *all* forces are relative" (Mach's emphasis).[5] Mach was attempting to disprove the existence of absolute space.

Such a disproof requires that there be no effect that can distinguish between a rotating laboratory in a fixed universe (the BR situation) and a fixed laboratory in a rotating universe (UR), but, more generally, or as Penrose has put it, that "physics should be defined entirely in terms of the relation of one body to another and that the very notion of a background space should be abandoned."[6] Yet, Mach should have concluded that a relational theory of space is inadequate and that there is no relativity of all motions by considering the following elementary examples. In IIA and IIB we consider interactions of two or more objects in a background universe and in IIC the interaction of a single object with the universe.

## II.A A relational theory of space is incomplete.

Example 1: There is no relativity of rotation for electromagnetism or gravitation.

Consider an inertial frame F with a spherically symmetric charge $Q$ of radius $R$ at the origin and a point charge $q$ at $x = 0, y = r, r > R,$ in an otherwise empty universe, or, in the alternative, in an universe filled with a perfectly homogeneous, isotropic, featureless, uncharged cold dust or "cold dark matter" with a mass-energy density and a rate of expansion equal to those at present (i.e. a universe identical to ours except for the absence of any object that can serve as a reference point). $Q$ has a marker $s$ on its surface at $x = 0, y = R$. Both classical physics and relativity allow us to distinguish four cases:

-- Case (a): $q$ revolves around $Q$ with angular velocity $\omega\mathbf{k}$. What is the magnetic field at $q$? One can calculate the electric field $\mathbf{E}$ due to $Q$ by using the law of Biot and Savart, with $\mathbf{v} = \omega r \mathbf{i}$:

$$\mathbf{B} = (v/c^2) \times \mathbf{E}. \qquad (1)$$

-- Case (b): $q$ is at rest but $Q$ is rotating with angular velocity $-\omega\mathbf{k}$. What is the magnetic field at $q$? Mach, as a believer in the principle of relativity of all motions, including the relativity of rotational motion, would maintain that cases (a) and (b) are equivalent and conclude that the answer is the same as before. This argument gives an incorrect answer because in Eq. (1) one uses *relative linear velocities*: in case (b) one must use the linear velocities of the infinitesimal charge elements of $Q$ relative to $q$ and, unlike case (a), one must know the radial dependence of the charge density in $Q$. (There is no magnetic field at $q$ if $Q$ is a rotating point charge.)

-- Case (c): $q$ and $s$ are aligned as $q$ revolves around $Q$ with angular velocity $\omega\mathbf{k}$ and $Q$ is rotating with angular velocity $\omega\mathbf{k}$. A relationist would maintain, wrongly, that there is no magnetic field acting on $q$ for any value of $\omega$ since there is no such field for $\omega = 0$.

-- Case (d): $Q$ revolves around $q$ with angular velocity $-\omega\mathbf{k}$ without rotating (as seen from $q$, the marker $s$ has an angular velocity $\omega\mathbf{k}$). The magnetic field at $q$ is the same as in case (a).

Cases (a), (b), (c), and (d) can also be distinguished from each other if we consider velocity- and acceleration-induced frequency shifts from sources on $q$ and $Q$, radiation emitted by $q$ and $Q$, or centrifugal forces in the sphere $Q$, with no centrifugal force only in case (a).

A relational theory of space cannot distinguish between cases (a), (b), (c), and (d): a relationist would maintain that there is only one observable, the angular velocity $\omega\mathbf{k}$ for the marker $s$, as seen by $q$, and thus that the four cases should give the same answer for **B,** for the Doppler shifts, and for the centrifugal force in $Q$.

Now, place a light source megaparsecs away. A relationist can distinguish between cases (a), (b), (c), and (d) (and also determine the value of $\omega$ in case (c)) by using Doppler shifts and the relationist must now explain how a light source so far away can affect the observed magnetic fields, Doppler shifts, and centrifugal forces. Thus we conclude that a relational theory of space is incomplete. Needless to say, $Q$ can be replaced by a bucket in all of the above examples.

The previous discussion addresses primarily electromagnetic forces, but SR teaches us that these are but the manifestations of Coulomb forces in the frame of reference where both the 'source' and the 'field' charges are in motion. Analogously SR and GR predict gravitomagnetic forces between two masses in a frame where the two masses are in motion. The conclusions from the examples we have considered apply to gravitomagnetic forces. The same is true for the remainder of this Section II.

**II.B There is no relativity of all motions**

Example 2: The laws of physics do not allow both linear and angular velocities to be relative.

Case (a): Given in an inertial frame F a point charge $Q$ at the origin and a point charge $q$ at $x = 0$, $y = a$, and let $Q$ and $q$ have velocities $\mathbf{v} = v\mathbf{i}$ and $\mathbf{v'} = v'\mathbf{i}$ respectively. What is the magnetic force between them? A classical physicist would ask whether F is at rest with respect to the ether. Mach believed in the ether, but he might have argued, given that he believed that all motion is relative, that the magnetic force depends only on the relative velocity $\mathbf{V} = (v'-v)\mathbf{i}$, with no magnetic force if $\mathbf{V} = 0$.

Case (b): For the charges above one can define $b$ so that $v/v' = b/(b+a)$ and choose $v$, $v'$, and $\omega$ so that $\mathbf{v} = \omega b \mathbf{i}$, $\mathbf{v'} = \omega(b + a)\mathbf{i}$. With these choices $\mathbf{v}$ and $\mathbf{v'}$ are instantaneous circular velocities of $Q$ and $q$ (in the reference frame F) as they revolve around the point $(0, -b)$ with angular velocity $\omega\mathbf{k}$. Because the two charges have no relative angular velocity, Mach, as a believer in the relativity of rotational motion, would conclude, wrongly, that there is no magnetic force between them.

Example 3. There is no relativity of accelerations or of centrifugal forces.

-- Case (a): Consider buckets A and B on the $x$ axis of an inertial frame F. Let A and B have velocities 0 and $v\mathbf{i}$, respectively. An observer on A cannot say s/he is "at rest." A and B are equivalent: we can choose a frame F' where A and B have velocities -$v$**i** and 0 respectively, and B would be just as much "at rest."

-- Case (b): Let $v$ =0 while A and B rotate with angular velocities 0 and +$\omega$**k** respectively. An observer on B's rim would say A has angular velocity -$\omega$**k**. Are observers here equivalent just as they were in the linear motion case (a)? NO! B exhibits centrifugal forces but A exhibits no centrifugal forces, even though it rotates with angular velocity -$\omega$**k** relative to B.

An analogous argument can be made for linearly accelerated motion: in F let A have no acceleration while B has acceleration $b$**i.** B could say that A has acceleration -$b$**i,** but the two descriptions are not equivalent because accelerometers on B yield non-zero readings, even in a frame where B's acceleration is zero. Thus we cannot transpose conclusions drawn from uniform linear motion to rotational motion or to linear accelerated motion unless we can show that such a transposition is valid.

-- Case (c): Let A and B both rotate with angular velocity $\omega$**k.** Having the same angular velocity is very different from having the same linear velocity. Observers on each bucket observe their own and the other bucket's rotational motion.

### *Comment on II.A and II.B: Should one not consider the motion of the universe?*

Relationists could object to our treatment in IIA and IIB in that we have kept the universe 'fixed' in all our examples. Regarding Examples 1(a) and 1(b) they would insist that the rotational equivalent of a point charge $q$ revolving around a charged sphere $Q$ is a rotation of $Q$ and of the whole universe. Schiff has indeed proposed that classical electrodynamics be reformulated so that these two situations would be equivalent and that in 1(c) the magnetic field would vanish for all values of $\omega$.[7] We do not believe that Schiff's treatment is adequate in that he does not address with respect to what frame could the universe be said to be rotating. Moreover if Example 1(a) is modified so that there are now several point charges $q_i$ each with its own $a$ and $\omega$ for revolution in its own orbital plane, there is no rotational equivalent that Schiff can propose, especially if all the $q_i$ are in fact extended rotating charge distributions. Rindler has shown that Schiff's model leads to further inconsistencies.[8]

In example 3 (c) there is no way one can explain what is observed by postulating that it is the universe that is rotating.

## II.C Mach intended his "relativity of rotation" to apply only to rotation of a single system relative to the universe and this theory violates SR and GR

In spite of his declarations, Mach did not use "relativity of all motions" in his work. He used a more restricted relativity of rotation, saying "try to fix Newton's bucket and rotate the heaven of fixed stars and then prove the absence of centrifugal forces."[5] (Mach did not specify how to rotate the universe.) Yet, because Mach was attempting to disprove the existence of absolute space, the invariance under rotation that Mach proposed cannot be limited to centrifugal forces. The invariance under rotations that Mach proposed must have been intended to be a universal law of nature. Since one cannot perform the rotating universe (UR) experiment we shall assume for sake of argument that all inertial forces, both centrifugal and Coriolis, are observed in Mach's fixed bucket in a UR situation. We will show that experiments, especially when SR and GR are taken into account, allow us to distinguish between UR and BR situations, and specifically, <u>between a non-rotating bucket in a non-rotating universe and a co-rotating bucket in a rotating universe, a distinction that goes against Mach's relational theory of space.</u>

-- <u>Parity Non-Conservation.</u> Nearly 250 years ago, Kant argued that the irreducible difference between the left and right hands is incompatible with a relational theory of space. Parity non-conservation in the weak interactions demonstrates such a difference at the sub-atomic level.[9]

-- <u>Electrodynamics.</u> Most importantly, the equivalence between BR and UR situations goes counter to classical electrodynamics and to SR. In Example 1 above, let $Q$ be a bucket filled with a charged liquid and $q$ a point charge in the universe: $q$ revolving around $Q$ (case (a)) is not the same as $q$ fixed while $Q$ is rotating (case (b)) because there is no relativity of rotation. Also, since transverse Doppler shifts depend on relative linear velocities, there are different transverse Doppler shifts in the BR and UR situations. These Doppler shifts allow us to distinguish between a non-rotating bucket in a non-rotating universe and a co-rotating bucket in a rotating universe.

In Example 2, let $Q$ and $q$ be charges at rest in the universe. There is no magnetic force between them under BR, but there is one as they revolve around B under UR. Also, there are different transverse Doppler shifts between these charges themselves and with respect to B in the BR and UR situations.

BD's response in Ref. 4 (Sec. IV) is the postulate that "[only] those who are non-accelerating with respect to distant stars are inertial observers [regardless of whether the whole universe of stars is rotating or not]." There is no empirical support for BD's postulate: there is no evidence that a large rotating galaxy acts as an inertial frame, not even in part. Defining a rotating universe as an inertial frame goes counter to contemporary relativistic physics, so we shall not assume BD's postulate is correct.

<u>Gravity and Inertia: the Lense-Thirring effect.</u>. Concerning inertia, BD assert that for Mach centrifugal forces are "gravitational," that is, produced by the action of mass, "*all* the masses of the universe," (our emphasis) upon another mass. Concerning the rotating liquid, BD also claim that "in the absence of all other particles in the universe … the surface of the [rotating] water would never become concave. Because this experiment cannot be performed, it is impossible to determine whether Mach or Newton is right." By insisting that one cannot determine whether Newton or Mach is right,

BD in fact are saying that Mach has *an alternate interpretation*, but not a theory. We maintain that one can prove Mach is wrong, that he has *a theory that is wrong.*

We will show in Sec. III that Mach's approach is not gravitational but *geometrical,* but here we will show that Mach's contention is counter to GR. The centrifugal force in a BR situation is proportional to the distance *r* from the axis and can appear in a rotating bucket in flat space-time, *with no centrifugal force outside the bucket*. Mach thus requires that with UR there be a centrifugal force near the axis of the rotating universe proportional to the distance *r* from the axis. Is it confined to the bucket and, if so, how? If it is gravitational, this force cannot be an artifact of the frame of reference and because it is *r*-proportional it constitutes a "tidal force." In GR a tidal force is a sufficient condition for space-time curvature, and curvature is invariant under rotations.[10] Because the curvature of space-time is different in UR and BR, the two situations are not equivalent. There are no solutions of the GR field equations that yield an *r*-dependent force at and near the origin such that the force vanishes abruptly at a certain radius, a force which changes whenever a different bucket is used. Finally, Mach maintained that the inertial rest mass $m_0$ of an object, for example a fluid element in a bucket, depends on the mass density of the universe. With SR, this mass density depends on the linear velocities of the stars relative to the bucket. Thus there is a different value for $m_0$ in the expression $\gamma m_0 \omega^2 r$ for the centrifugal force on the fluid element in BR and UR situations.

BD note that GR predicts an *r*-proportional centrifugal force in the equatorial plane of a rotating massive shell ( the 'Lense-Thirring effect') and claim that this centrifugal force is a Machian effect; that is, the rotating shell corresponds to a UR situation which, following Mach, would in turn be equivalent to a BR situation. But the centrifugal force in the rotating shell is not equivalent to that in a BR situation because (a) it is not limited to the inside of a bucket and (b) it produces space-time curvature.

BD neglect other effects in a rotating mass shell: (1) a *z*-dependent axially directed *centripetal* force near the equator that is twice as large as the centrifugal force and for which there is no BR equivalent, (2) transverse Doppler shifts for sources on the shell when seen from the inside, and (3) velocity-dependent Coriolis-type forces that produce "frame dragging" throughout the inside of the shell Both the centrifugal and the centripetal forces in the rotating shell are attributable to the SR increase in the mass near the equator because of its larger linear velocity.[11] The same special-relativistic mass anisotropy and centripetal/centrifugal force combination should appear under UR. Also, in their Sec. III, BD deny the presence of frame dragging in the UR case.

The arguments in the previous three paragraphs show that by such means as transverse Doppler shifts, curvature of space-time, and a centripetal acceleration along the axis of rotation, SR and GR distinguish between a non-rotating bucket in a non-rotating universe and a co-rotating bucket in a rotating universe, distinctions that are impossible in a relational theory of space.

Although for Mach it is meaningless to speak of an empty universe, or of an isolated object in an otherwise empty universe, GR offers predictions for an empty universe, introducing a "cosmological constant" aka "dark energy" (recent measurements indicate that such a universe expands.)[12] GR also has solutions for various isolated objects in an otherwise empty universe, most notably the Kerr-Newman solution for a rotating charged object where rotation-induced inertial and magnetic effects are predicted.

Einstein himself eventually abandoned the search for a Machian formulation of GR.[13]

A possible construal of the Mach-BD position is that the universe, whether rotating or not, constitutes a plenum (we prefer not to use the word 'ether') such that all motions are relative to that plenum.

## II.D GR does not allow for a 'Machian' rotating universe.

While there are several rotating relativistic cosmological models, all models that are homogeneous in space and time are unphysical.[14] Godel's model has received the most attention and it has been shown that it does not fulfill Machian expectations. Godel's energy-momentum tensor is the same as Einstein's in his static universe model but the geometry of space-time is different, thus, contrary to Mach's teaching, the distribution of matter in the universe does not determine the geometry of spacetime.[15] Also, the Godel spacetime features closed time-like curves making it acausal.[15] It is only through the imposition of a causality condition that GR yields a unique spacetime.

## III. Mach's "relativity of rotation" is internally inconsistent even when it is restricted to rotations with respect to the universe.

### III.A There cannot be a fixed bucket in a rotating universe.

In Ref. 3 we claimed that a fixed bucket in a rotating universe is inconsistent with Mach's relativity of rotation. We considered a friction-free bucket, the universe (represented by a star S), and a friction-free pendulum with a friction-free bob suspended above the bucket and swinging with the same frequency as the bucket's rotation. The bucket has a V painted on the bottom, the star a bright spot pointing to the bucket marked by S>, and the orientation of the bob of the pendulum is designated by Δ. If the system starts with the initial orientation

S>    Δ    V,

then half a period later the relative positions and orientations are different in the BR and UR cases: in the UR case, the bucket is never between the star and the bob. Only if one assumes that the rotation of the universe induces rotation of the pendulum plane ("frame-dragging") and that this frame-dragging also induces rotation of the pendulum bob does one obtain identical situations in the BR and UR cases. Then we argued that if a rotation is induced in the bob, a rotation must also be induced in the bucket. Thus we concluded that Mach's ideas require frame-dragging, but that frame-dragging implies that there cannot be a fixed bucket in a rotating universe. A detailed analysis in GR of the effects inside a rotating massive shell shows that there cannot be a "fixed" object inside such a shell and thus supports our conclusion.[16]

We also pointed out that to have UR/BR equivalence when the star S revolves around the bucket, S must rotate with respect to the bucket from an S> orientation to an <S orientation so that a Sagnac experiment on S would detect that rotation; we also questioned the necessity of this rotation for a gravitation-based theory of inertia.

BD's response is that in Ref. 3 we erred in not assuming that the pendulum plane and the bob must necessarily rotate in the UR situation: "anything within the universe will move with it. The bucket is kept fixed (by applying external forces just like those needed to rotate the bucket in a static universe) and is not allowed to move with the universe." BD deny that there is frame-dragging. Also, they deny that there is simultaneous rotation of the star S as it revolves around B (while its bright spot always points towards B) and thus they deny a Sagnac effect on S, but we do not understand the basis for these claims.

There is no need to apply a force to keep an object, for example, a galaxy, rotating in a BR situation. BD's position that a mechanism is needed to keep B rotating under BR and fixed under UR is puzzling. What keeps a galaxy rotating? Does not every stable gravitationally-bound system rotate? Our condition that B be mechanically decoupled from the universe under UR is necessary for BR/UR equivalence. And this equivalence requires simultaneous rotation and revolution of a star S as the universe rotates, just as the Moon rotates as it revolves around the Earth while always presenting the same face to the Earth (see Sec. IV).

### III.B A universe rotating as a rigid body leads to further inconsistencies.

According to Ref. 4, Mach's relativity of rotation requires that the universe rotate as a rigid body and that the "universe" include everything except the object taken to be the bucket that rotates under BR. Yet BD claim that centrifugal forces are "gravitational," that is, produced by the action of mass, "*all* the masses of the universe." (our emphasis. See Sec. II in Ref. 4). Thus for a molecule where the atoms can be separated into two groups A and B such that B can rotate with respect to A, BD's position in countering the reasoning presented in IIIA is that the centrifugal force on the atoms in group B can be explained only if group A is included in the rotating universe while group B is fixed. How can the mass of the A atoms be the gravitational *sine qua non* cause for the inertia of the B atoms? In typical gravitational theories a small increment in mass produces only a small change in any gravitational effect outside the mass, even if a horizon is formed because of that increment. In Ref. 3 we could have had the pendulum submerged in a bucket filled with a low-viscosity fluid. BD's position is that an invisible hand must make that pendulum rotate while the bucket and the fluid are fixed. Thus BD rely on a *geometrical point-to-point identity between the UR and BR situations rather than on masses revolving around B exerting a gravitation-type centrifugal force in B*. In other words, *Mach's relativity of rotation applies only to objects that are as structure-free as an electron.*

Also, the condition that the rotating universe include everything but the bucket in question cannot be applied consistently:

Case (a): Consider two coaxial buckets A and B that manifest centrifugal forces under BR when both are rotated. Keep A and B uncoupled from the universe and rotate the universe: according to Mach, both should manifest centrifugal forces.

Case (b): Reverse the sequence, doing the UR experiment first, with both A and B fixed and uncoupled from the universe. According to Mach, both manifest centrifugal forces again. Now stop the universe and rotate only B, leaving A fixed. B exhibits centrifugal forces in spite of the fact that everything did not rotate while B was kept fixed. (A did not rotate with the universe in the second experiment.) Thus it is not necessary that A be attached to the universe for Mach's experiment to work out for B. Similarly, contrary to Ref. 4, the pendulum in Ref. 3 that is swinging above the bucket need not be attached to the universe as the universe rotates around the bucket.

## IV. Differentiating rotation of the universe from rotation of the bucket

In Ref. 4, Sec. IV, BD claim that "an observer rigidly attached to the bucket, by looking at the motion of … distant stars, he/she will conclude that either the bucket is rotating and the universe is static or vice versa (similar to what a traveler experiences sitting on a train). So the observer can describe his/her observation from two different perspectives, BR and UR, and, according to Mach both observations are equally true." This statement contradicts SR: the UR observer can determine: (a) the relative angular velocity $\omega$ between the bucket and the universe by measuring the time between transits of a specific star; (b) the distance $d$ to that star by the parallax method; and (c) the relative linear velocity $v$ of the star with respect to the bucket rim (radius $a$) by means such as the transverse Doppler shift or stellar aberration. The velocity $v$ would be $\omega a$ for BR, $-\omega d$ for UR, and an intermediate value if both the universe and the bucket are rotating. If $|v| > \omega a$, the observer must conclude that the star is revolving around the bucket. (In Ref. 3 we noted that Mach requires that the concavity of the liquid and the measurement of the star's transit time yield the same value for $\omega$ while Newton does not.)

More generally, see Sec. II, BD's equivalence between linear and rotational motion is not correct. If the bucket rim and the distant star S are in relative uniform rectilinear motion, Doppler shifts and stellar aberration would be the same no matter whether an observer is located on S or on B. Also, if two objects have different uniform velocities, they will cross each other innumerable times if they travel on a circular track – but no more than once on a rectilinear track.

Given our ability to distinguish between BR and UR situations, all of BD's arguments in their Sec. IV are unavailing, especially those pertaining to the Sagnac effect. We examine some of them in detail to bring out some interesting features.

### IV.A Rotationally induced electromagnetic fields

In Ref. 3 we questioned the equivalence of BR/UR as follows. A magnetic field is produced when an electrically harged liquid is kept in a rotating bucket and an electric field is induced in a wire above the liquid, rotating with the bucket and perpendicular to the axis of the bucket. On the other hand, these electric and magnetic fields cannot exist under UR. BD respond that the calculation of the magnetic field caused by a rotating liquid "is performed with respect to an inertial observer," who, Mach defines, is "not accelerating with respect to the distant stars. For both UR and BR, the liquid is rotating with respect to the observer tied with the universe. Hence the same derivation is applicable for both the UR and BR cases."

BD are not correct. Let the bucket have a charge $Q$ and consider a point charge $q$ in the universe (this case is a variant of Example 1 in our Sec. II). BR corresponds to case (b) with a magnetic force on $q$ depending on the details of the charge distribution in $Q$. UR corresponds to case (a) with the magnetic force on $q$ depending only on the magnitude of $Q$. There is no BR/UR equivalence.

## IV.B Radiation from a point charge on the bucket

In Ref. 3 we argued that, under BR, a charge on the bucket rim would radiate and the bucket would lose energy, while under UR this would not happen. BD respond that the relative acceleration between charges on the bucket and in the universe is the same under BR and UR.

Their argument is an invocation of Shiff's postulated electrodynamics. BD have to show that a rotating universe causes a charge in a fixed bucket to emit photons and that the photons seen by observers in both the bucket and the universe depend only on the charge's distance from the axis and the universe's rotation rate. Under UR, why would observers in a fixed bucket see photons emitted by charges on that bucket? The equality of the relative accelerations under BR and UR is an assumption for BD, but it is not presently accepted to be correct physics: the acceleration of the charges is $\omega^2 r$, with $r$ many orders of magnitude larger for UR than for BR.

## IV.C The Sagnac effect in a rotating universe

BD state correctly that the "Sagnac effect can be simply described as the difference in round trip travel time of co-rotating and counter rotating light rays from the viewpoint of an observer on a rotating platform." This difference in time leads to a shift in the interference fringes when two oppositely rotating beams are superimposed. BD point out that there are many competing explanations for the Sagnac effect, but they do not deny that it occurs in every rotating system.

BD deny that a Sagnac effect would be observed on a star that is part of the rotating universe. A star S participating in the rotation of the universe rotates as it revolves around the bucket B because the star always presents the same face to B. If the distances from B of two diametrically opposite points 1, 2 on S are denoted by $S_1$ and $S_2$, then these distances remain the same during UR. But if 1 and 2 are collinear with B the velocities $V_1 = \omega S_1$ and $V_2 = \omega S_2$ are different. Thus S rotates around the point midway between 1 and 2, just as the moon rotates as it orbits the earth. A Sagnac effect experiment on the star S can detect this rotation.

BD argue that an inertial observer would observe the effect and they argue that, according to Mach, an observer who is at rest with respect to a distant fixed star is an inertial observer. An observer on the bucket axis and rotating with the universe will observe the Sagnac effect on the star. An observer on the bucket axis at rest with the bucket will also observe the effect. A Sagnac experiment can also be performed in the rotating universe itself, with a path linking galaxies surrounding the fixed bucket.

### IV.D Superluminal velocities as the universe rotates

In the context of classical physics, i.e. the physics Mach accepted, we argued in Ref. 3 that in the BR case the tangential velocity $V$ of a point on the rim of a bucket of radius $a$ is $\omega a$, whereas for UR the tangential velocity $v$ of a star S at a distance $r$ from the axis of rotation would be $\omega r$. Because $r$ could be very large for a distant star, $v$ could easily exceed $c$, that is, $v$ could be superluminal. These velocities are relative to the center of the bucket, which is also the center of the universe in both BR and UR. We have no relativistic model for a rotating universe except when the frequency of rotation $\omega$ is so small that $v=\omega R<c$, where $R$ is any cosmological distance. On the other hand, for a terrestrial bucket with $\omega =1$, having the universe rotate instead will have the moon have a velocity larger than $c$. Thus Mach's "rotate the universe" is an empty proposal.

Whatever $\omega$ may happen to be one will have $V = \omega a \ll \omega R = v$. Relativity teaches that an observer at the center can measure both $v$ and $V$ by means of transverse Doppler shifts, for which the observer's rotation is irrelevant. There is no doubt that observed Doppler shifts are much larger with UR than with BR; $R$ can be $c$ times the age of the universe, and a bucket rotation period as slow as 1/6 the age of the universe will generate superluminal velocities under UR. Therefore *Mach's relativity of rotation applies only to buckets with infinitesimal angular velocities.*

### IV.E Bucket/water interactions

In Ref. 3 we further argued that in the BR scenario water droplets fly off tangentially at a constant velocity when they leak out from the bucket, while they would fly off radially in the UR case. Under BR, if a droplet leaks out at $t = 0$ with coordinates $(0, a)$ from a bucket of radius $a$ and angular velocity $\omega \mathbf{k}$, its position $\mathbf{P}$ and velocity $\mathbf{V}$ with respect to the bucket point from which it leaked are

$$\mathbf{P} = (a\omega t - a\sin\omega t)\mathbf{i} + (a - a\cos\omega t)\mathbf{j} \qquad (2)$$

$$\mathbf{V} = (a\omega - a\omega\cos\omega t)\mathbf{i} + a\omega\sin\omega t\,\mathbf{j}. \qquad (3)$$

BD cannot obtain the same $\mathbf{P}$ and $\mathbf{V}$ under UR, which is required if Mach is correct.

BD assume that under UR, with the bucket fixed, the droplet is ejected with no centrifugal force (why does the gravitational force of the rotating universe, which BD claim causes the centrifugal force, stop at the rim?) and that the droplet has velocity $-a\omega\mathbf{i}$ with respect to the rotating universe. But this velocity is the same with respect to the universe as that of the bucket rim, and thus the droplet should stay with the bucket. There is no ejection! BD claim that once out of the bucket the droplet becomes part of the universe and rotates with it so that the net motion is a spiral, as seen from the bucket. This spiral is not at all what we observe under BR.

Rather than trying to obtain Eqs. (2) and (3) for **P** and **V** under UR, BD consider displacements only in the *x* direction and only when $\omega t = \pi$, the only time when $P_x$ is the same for BR for UR. They do not consider **V** at all, nor the **j** direction displacements. How can the $-a\sin\omega t\,\mathbf{i} - a\cos\omega t\,\mathbf{j}$ terms in **P** and the $-a\omega\cos\omega t\,\mathbf{i} + a\omega\sin\omega t\,\mathbf{j}$ terms in **V** be duplicated if the bucket is fixed? Only if $a = 0$! Thus *Mach's relativity of rotation applies only to point-like buckets*. But, can a point-like object rotate?

### IV.F Multiple buckets

In Ref.3 we showed that Mach's relational approach leads to a very awkward treatment of a system containing several buckets. There is no doubt that given an arbitrary number *n* of buckets rotating around arbitrary axes, an observer on bucket *i* can describe his/her observations in terms of bucket *i* being at rest and non-rotating and all the other buckets revolving around *i* and rotating around their axes. But Mach admonished us that we are given the world only once.[5] In light of that admonition, must we not choose a description and a theory that treats all *n* buckets equally? Newton did that, Mach did not.

### V. The "equal actuality" of the Heliocentric and Geocentric Systems

BD maintain that "a corollary of relativity of rotation is that the rotation of earth around the Sun is equivalent to the rotation of Sun around the Earth." [We suggest that one speak of the *revolution* of the earth around the Sun and vice versa.] We have already shown that there is no relativity of rotation in general (see Sec. II.A), and BD have not shown that Mach's relativity of rotation applies here. The equivalence proposed by BD is not Mach's BR/UR equivalence: the latter would have the earth, as an analog to the bucket rim, be fixed with respect to the Sun while the universe rotates around the Sun.

BD's arguments focus on the notion that an observer O on any planet $X$ in the solar system can describe his/her observations in terms of $X$ being fixed and all other objects in the universe revolving around $X$ and that we can formulate the laws of physics so as to "explain" O's observations. (BD do not explain how to treat simultaneous observations on planets $X1, X2, X3$ etc…, especially if they belong to different stellar systems.) But the possibility of different choices of reference systems is beside the point: Mach was not interested in equivalent descriptions. He claimed that the Copernican and the Ptolemaic systems were *equally actual*.[5] BD do not explain why Mach wrongly used the term "Ptolemaic system" which they concede Galileo had falsified, rather than the Brahean system, which is more plausible. We think that "Ptolemaic" and "equally actual" were chosen because of pamphleteering bravura. Mach proclaimed "We are given [the world] only once."[5] This last argument suggests that we adopt a theory applicable for observations made simultaneously from all planets in all planetary systems. Newton did that, Mach did not.

Mach and other proponents of the equivalence between the Ptolemaic-Brahean and the Copernican systems limit themselves to the observations available in the 16th century, arguing that these observations (the angular positions of the stars and planets) can validate both systems. This approach has history stopping before the phases of Venus and stellar and planetary aberrations, parallaxes, and Doppler shifts were observed. Mach ignored these observations and BD do not explain why he did so, nor do they explain how a Ptolemaic-Brahean system can accommodate them.

Also, we should ask what an observer outside the solar system would observe about the motions of the solar planets. We know much more astronomy than Mach. We have thousands of observations of clusters of galaxies, individual galaxies, double-star systems, and planetary systems beyond the Sun. There is not a single case where an object of a given mass revolves around an object of a lesser mass. What we see is both objects revolve around their system's center of mass, located near the heavier mass. Why should the solar system be different? BD provide no reason.  Also, the cosmic microwave background shows a 365-day period in the frequency of the radiation observed in the direction of a given galaxy in the ecliptic plane.  This indicates that the earth moves through the background.

Also, in a geocentric system even the closest stars have superluminal velocities. Thus there is no reason why a geocentric system should be considered.

**VI. Discussion**

Mach's ideas have intuitive appeal. It makes sense to say that a Foucault pendulum at the North Pole keeps on swinging toward star S because it was set in motion in that direction and the combined forces from all the masses in the universe make it maintain that direction. This explanation of the observations may just be a convenient way of speaking with S being a marker for a point in space and "the combined forces" a question-begging assertion. Although Einstein was much influenced by Mach when he began developing GR, the latter is even more in conflict with Mach's ideas than was Newtonian physics. With Newton, space acts but is never acted upon. With Einstein, the motion of bodies affects the structure of space-time, especially when work must be done to accelerate these bodies. Einstein provides us with a plethora of means for distinguishing between a bucket rotating in a fixed universe and a bucket fixed in a rotating universe and, specifically, between a non-rotating bucket in a non-rotating universe and a co-rotating bucket in a rotating universe, a distinction that goes against Mach's relational theory of space. In Ref. 3 and here we have presented only a glimpse of the many ways in which classical and relativistic physics contradict Mach. Readers should consult Ref. 9 and the references therein for a more thorough presentation of the issues.

NOTES